\documentclass[final,5p,times,twocolumn]{elsarticle}

\usepackage{graphicx}
\graphicspath{{Figures/}}
\usepackage{amsmath,amssymb}

\biboptions{numbers,sort&compress}

\journal{Computational materials Science}

\begin{document}

\begin{frontmatter}


\title{Determining phase transition using potential energy distribution and surface energy of Pd nanoparticles}



\author[af1]{Maryam Azadeh\fnref{af3}}
\author[af2]{Movaffaq Kateb}
\author[af1]{Pirooz Marashi\corref{cor1}}

\address[af1]{Department of Mining and Metallurgical Engineering, Amirkabir University of Technology, Tehran, Iran}
\address[af2]{Science Institute, University of Iceland, Dunhaga 3, IS-107 Reykjavik, Iceland}

\fntext[af3]{Present address: School of Metallurgy and Materials Engineering, University of Tehran, Iran}
\cortext[cor1]{Corresponding author: pmarashi@aut.ac.ir}

\begin{abstract}
Molecular dynamics simulation is employed to understand the thermodynamic behavior of cuboctahedron (cub) and icosahedron (ico) nanoparticles with 2 -- 20 number of shells (55 -- 28741 atoms). The embedded atom method was used to describe the interatomic potential. Conventional melting criteria such as potential energy and specific heat capacity ($C_p$) caloric curves as well as structure analysis by radial distribution function ($G(r)$) and common neighbor analysis (CNA) were utilized simultaneously to provide a comprehensive picture of the melting process. It is shown that the potential energy distribution and surface energy ($\gamma_p$) proposed here are holding several advantages over previous criteria. In particular, potential energy distribution can distinguish between interior and surface atoms and even corner, edge and plane atoms at the surface. While $G(r)$ and CNA are not surface sensitive methods and cannot distinguish between surface melting and an allotropic transition. It is also shown that allotropic change appears more clearly in $C_p$ and $\gamma_p$ rather than potential energy. However, determining accurate $C_p$ requires enough sampling to be averaged. Finally, a few issues in the current methods for determining $\gamma_p$ were discussed and a simple method based on available models was proposed which, independent of estimation of the surface area, predicts the correct temperature and size-dependent trend in agreement with Guggenheim-Katayama and Tolman's models, respectively.

\end{abstract}

\begin{keyword}
Nanoparticle \sep Melting \sep Allotropic transition \sep Potential energy distribution \sep Surface energy


\end{keyword}

\end{frontmatter}


\section{Introduction}




Nanoparticles are known as the most unstable structures among different nanosolids due to the higher surface to volume ratio \citep{Schmidt98}. Thus, it is crucial to develop a quantitative understanding of their thermodynamic stability for practical applications specifically when the thermal instability can be considered as a failure at the elevated temperatures. For instance, it has been shown that Pd clusters might experience a solid state transition \citep{kateb2018} below their melting temperature ($T_{mp}$) which has been misinterpreted as surface melting previously \citep{Pan05}. This indicates, importance of developing proper tools or methods for determine quantitatively correct values. In this regard, the main attention has been brought to Monte Carlo (MC) \citep{Westergren03} and molecular dynamics (MD) \citep{Baletto02,Schebarchov06} simulation which have proven to be an excellent tool for understanding the stability and melting behavior of nanoparticles.

Several methods are used in MD simulations to identify the melting process based on atomic specifics. The first criterion proposed by \citet{Lindemann10}, stating the melting of crystals occurs when the average amplitude of atomic vibrations, is higher than the threshold value. The global Lindemann index ($\delta_L$) is a system average of atomic quantities which shows a linear increase with the temperature increment in solid-state regime and a step change due to the melting. However, most of vibrations of the surface atoms in the small clusters, which have more degree of freedom, assumed as melting behavior by this model \citep{Alavi06,Zhang13}. This is a serious issue since it may lead to misinterpretation of the surface melting.

We have recently shown that a combination of various criteria such as caloric curves and structure analysis is required to study the phenomenological melting of nanoparticles \citep{kateb2018}. In this view, the advantage of a more recent structure characterization method of common neighbor analysis (CNA) over radial distribution function ($G(r)$) has been discussed. It has been shown that CNA facilitates observation of an allotropic change which could be confused with the surface melting using $G(r)$. On the other hand, CNA treats the surface atoms as a disordered structure due to the lack of symmetry at the cluster surface. While separate $G(r)$ can be defined for each cluster shell including surface atoms allowing to some extent study surface phenomena. This difference arises from the fact that $G(r)$ was originally developed to determine number of nearest neighbors while CNA is based on bond angles between \emph{1st} nearest neighbors. 

A more conventional method for determining a transition temperature is through caloric curves such as an isotherm change in the cluster potential energy ($U_p$) \citep{Shim02,Zhao01} or a peak in specific heat capacity ($C_p$) \citep{Qi01}. Unlike structure analysis, caloric curves can not provide any information on the mechanism of melting or phase transition. For instance, it has been shown that cuboctahedron (cub) clusters are melting almost uniformly by nucleation of melt at (100) planes of surface and its propagation inward, while icosahedron (ico) nanoparticles are melting diagonally starting from a corner \citep{kateb2018}. It has also been shown that cub to ico transformation is achieved through a \emph{transitional disordered state} which can be carefully monitored by CNA \citep{Schebarchov06,kateb2018} while it only appears as a minor peak in $C_p$ or a small step change in $U_p$ which was ignored in the previous studies \citep{Qi01,Wang03,Pan05}. However, one can easily obtain the melting enthalpy of clusters ($\Delta H_{mp}$) from caloric curves. 

Another important criterion is the surface energy of clusters ($\gamma_p$) which plays an important role in their phase transformation \citep{Fischer08}. It is usually defined as the work of cutting a cluster out of the bulk material per unit area \citep{Buff51}. In contrast with CNA, $\gamma_p$ is very sensitive to the arrangement of surface atoms allowing to determine the melting temperature and solid state transitions. Regardless of this potential, utilizing $\gamma_p$ for detecting a phase transition is barely studied \citep{myasnichenko2016}. For instance, it has been shown experimentally that (110) planes melt completely with surface melting mechanism and (100) planes show partial surface melting while there is no surface melting for (111) surfaces \citep{Mei07}. This trend can be explained by the difference in surface energies, i.e.\ $\gamma_{(110)}>\gamma_{(100)}>\gamma_{(111)}$ \citep{Sankaranarayanan05}, which makes (111) planes more resistant against surface melting. 


In the present study we demonstrate utilizing potential energy distribution and surface energy for determining solid state transition as well as melting temperature. The result are compared to conventional criteria such as caloric curves, $G(r)$ and CNA. 

\section{MD Simulation}
\subsection{Simulation procedures}
MD simulations were performed by solving Newton's equation of motion \citep{Allen89} using large-scale atomistic/molecular massively parallel simulator (LAMMPS) \citep{Plimpton95} open source code, version 22 August 2018 (available at http://lammps.sandia.gov/).
The embedded atom method (EAM) \citep{Foiles86,Daw93} was utilized to describe the interatomic potential between Pd atoms. Eq.~(\ref{eq:eam}) represents the formulation of EAM potential:

\begin{equation}
	E_i=F_i \left[\sum_{i\neq j}\rho_{ij}(r_{ij})\right]+\frac{1}{2}\sum_{i\neq j}U_{ij}(r_{ij})
    \label{eq:eam}
\end{equation}

where $E_i$ and $F_i$ are cohesive and embedding energies of atom $i$, respectively. $\rho_{ij}(r_{ij})$ is the electron density of $j$ atoms located around the $i$ atom at the distance $r_{ij}$. Clearly, $F_i$ is a many body interaction term while $U_{ij}$ takes the pair interaction into account.

The EAM potential is extensively used for describing solid characteristics such as cohesive energy and elastic constant \citep{Foiles89} as well as metals melting temperature \citep{Foiles89,Foiles85}. Moreover, it is reliable in determining the transitional properties especially the heat of fusion and heat capacities above the room-temperature \citep{Mei91}. The EAM has also been verified for quantitatively correct description of such nanoscale systems namely surface energy and geometry of low index surfaces \citep{Foiles86,Daw84}. We have recently showed that the structure factor of molten Pd obtained by EAM potential is in close agreement to that of tight-binding and experiment \citep{kateb2018}.

The time integration of the equation of the motion was performed using the Verlet algorithm \citep{verlet67,Kateb12} with a timestep of 3~fs. The temperature was controlled by Nose-Hoover thermostat with damping time of 30~fs. These conditions are designed to generate positions and velocities sampled from canonical (NVT) ensemble. The initial velocities of the atoms were defined randomly from a Gaussian distribution at the 300~K and system was relaxed for 300 ps in the NVT ensemble. The melting simulations were performed by starting at 300~K, and then the temperature was elevated at a heating rate of 1.4$\times$10$^{12}$~K/s.

\subsection{Cluster preparation}

The Pd nanoparticles were considered to be in the cub and ico forms as found in experimental characterizations \citep{Jose01}. It is worth noting that in theoretical modeling routine, several structures and apparent shapes are considered. However those structure transform to more spherical shapes, ico and cub at elevated temperature as reported previously \citep{Wen09}. For different sizes of given clusters, cub and ico were made based on the so called magic number ($N_t$), total number of atoms, which is described as the function of the shell number ($n$) \citep{poole03}:

\begin{equation}
	N_t=\frac{1}{3}[10n^3+15n^2+11n+3]
\end{equation}
where $n=0$ denotes a monoatomic system and $n>1$ defines the full shell clusters.

Here, clusters with sizes below 12~nm including clusters of $n=2-20$ ($N_t=55-28741$ atoms) were chosen. 

\subsection{Visualization}

Version 2.9.0 of the open visualization tool (OVITO) package were used to generate atomistic illustrations (available at http://ovito.org/) \citep{Stukowski09}.

\section{Results and discussion}
\subsection{Relaxation}

\begin{figure}
	\centering
    \includegraphics[width=1\linewidth]{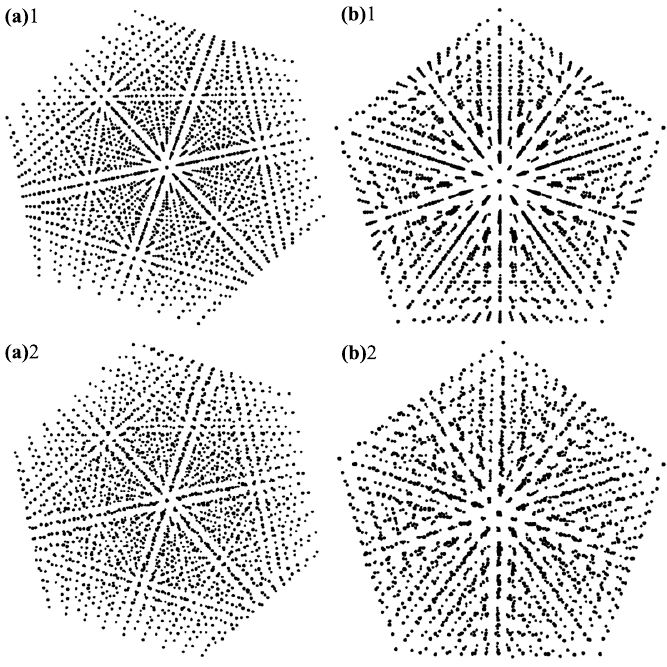}
    \caption{Atomistic view of (a) 8-cub and (b) 8-ico clusters before (a,b)1 and after (a,b)2 relaxation at 300~K, for 300~ps. Perspective view clearly shows there is no change in the symmetry and shape of clusterss.}
    \label{fig:relax}
\end{figure}

At the beginning, each nanoparticle was relaxed at room temperature as discussed in section 2.1 to minimize the potential energy of the entire system. Figure \ref{fig:relax} shows the atoms coordination, before and after relaxation for 8-cub and 8-ico cluster. The perspective view clearly shows that there is no change in the shape and symmetry of the particle, indicating current EAM potential can successfully model the stable shapes of Pd nanoparticles. The exceptions were 2-cub and 4-cub cluster that present a cub to ico transformation during the relaxation. This is in agreement with the previous result using a non-dynamic minimum energy calculation \citep{Baletto02}. It is worth mentioning that, we still call them 2-cub and 4-cub in the following.

\subsection{Melting criteria}
\subsubsection{Caloric curves}
\begin{figure}
	\centering
    \includegraphics[width=1\linewidth]{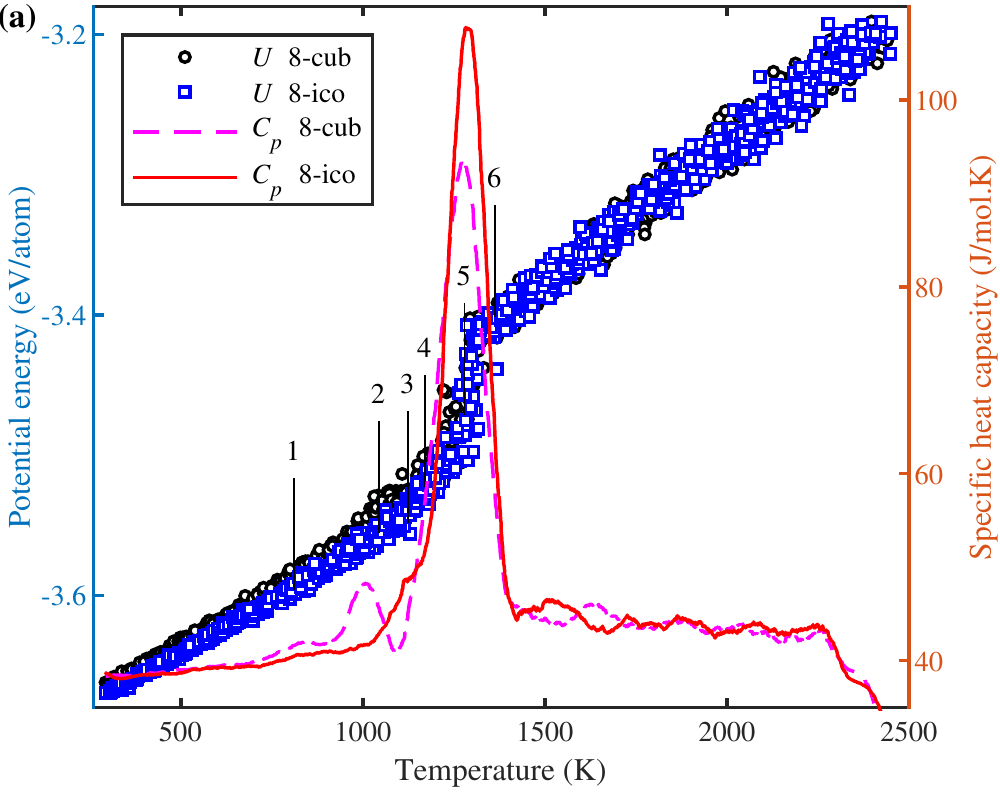}
	\includegraphics[width=1\linewidth]{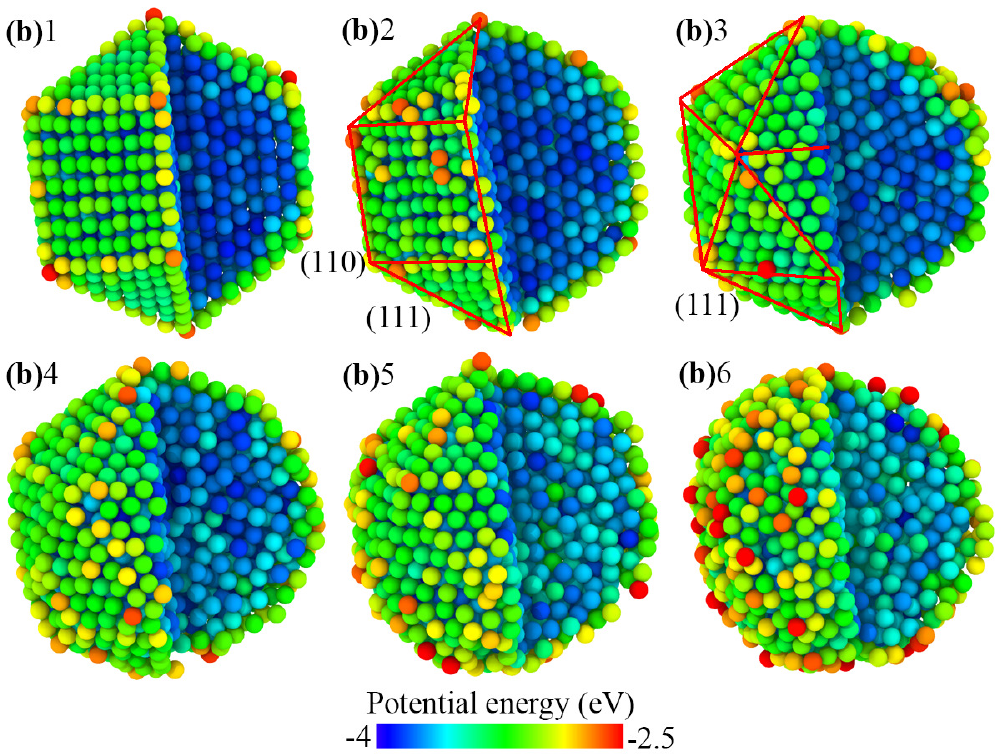}
    \caption{(a) Variation of $U_p$ and $C_p$ with temperature for 8-cub and 8-ico clusters and (b) snapshots of 8-cub cluster corresponding to points 1 -- 6 indicated in the (a).}
    \label{fig:caloric}
\end{figure}

Figure \ref{fig:caloric} illustrates the variation of $U$ and $C_p$ with $T$ for the 8-cub cluster (including 2057 Pd atoms) and corresponding snapshots. $U$ was determined by averaging over potential energy of entire atoms in the cluster and $C_p(T)=\partial\langle U_p\rangle_T/\partial T+\frac{3}{2}R$ with $R$ being universal gas constant \citep{Qi01}. The figure also contains the results of 8-ico cluster for comparison. The caloric curves present a typical melting behavior, i.e.\ an isotherm transition of $U$ corresponding to the main peak in the $C_p$ due to the latent heat of fusion. $T_{mp}$ equal to 1274 and 1288~K were determined using $C_p$ for the 8-cub and 8-ico clusters, respectively. In the caloric curve of 8-cub cluster, there is a step change at $\sim$1070~K corresponding to the local minimum in $C_p$. Such a behavior has been observed before however barely discussed in most cases \citep{Pan05,Baletto02}. For instance, \citet{Pan05} interpreted the $C_p$ minor peak as the surface melting \citep{Pan05}. However, they reported the surface melting for both cub and ico clusters without such a minor peak for the ico. Thus it is highly unlikely that local minimum or minor peak in $C_p$ are associated with surface melting. Apparently \citet{Zhang10} were the first who noticed such a step change in the caloric curve and attributed it to a solid-state transition using CNA. Utilization of heat capacity has been also demonstrated for a more complex transition at elevated temperature in Pt-Pd alloy \citep{chepkasov2017,chepkasov2018}. In this case, a visible change in the nanoparticle shape and number of surface atoms has been shown to be associated with the heat capacity fluctuations. Snapshots (b)1 -- (b)3 of figure~\ref{fig:caloric}, indicate no surface melting but slightly rounding in the corners due to surface diffusion. However, It can be clearly seen that (111) planes enclosed by triangles in (b)3 appear at the expense of vanishing (110) plane indicated by a square in (b)2. Snapshots (b)4 and (b)5 were taken close to the $T_{mp}$ that indicate a diagonal melting, i.e.\ faceted (111) planes on the top corner and rounded corner at the bottom. This incident might be associated with the partitioned structure of ico which retards the growth of the liquid phase. Finally, (b)6 snapshot shows the complete melting of the cluster. 

\subsubsection{Radial distribution function}
The first and simplest structure analysis is offered by the radial distribution function, $G(r)$. It describes how the atom number density ($\rho$) varies as a function of distance from a reference atom ($r$).

\begin{equation}
	G(r)=4\pi r^2\rho dr
\end{equation}
where $dr$ is the bin size or thickness of the spherical shell in which number of atoms is counted.

\begin{figure}
	\centering
    \includegraphics[width=1\linewidth]{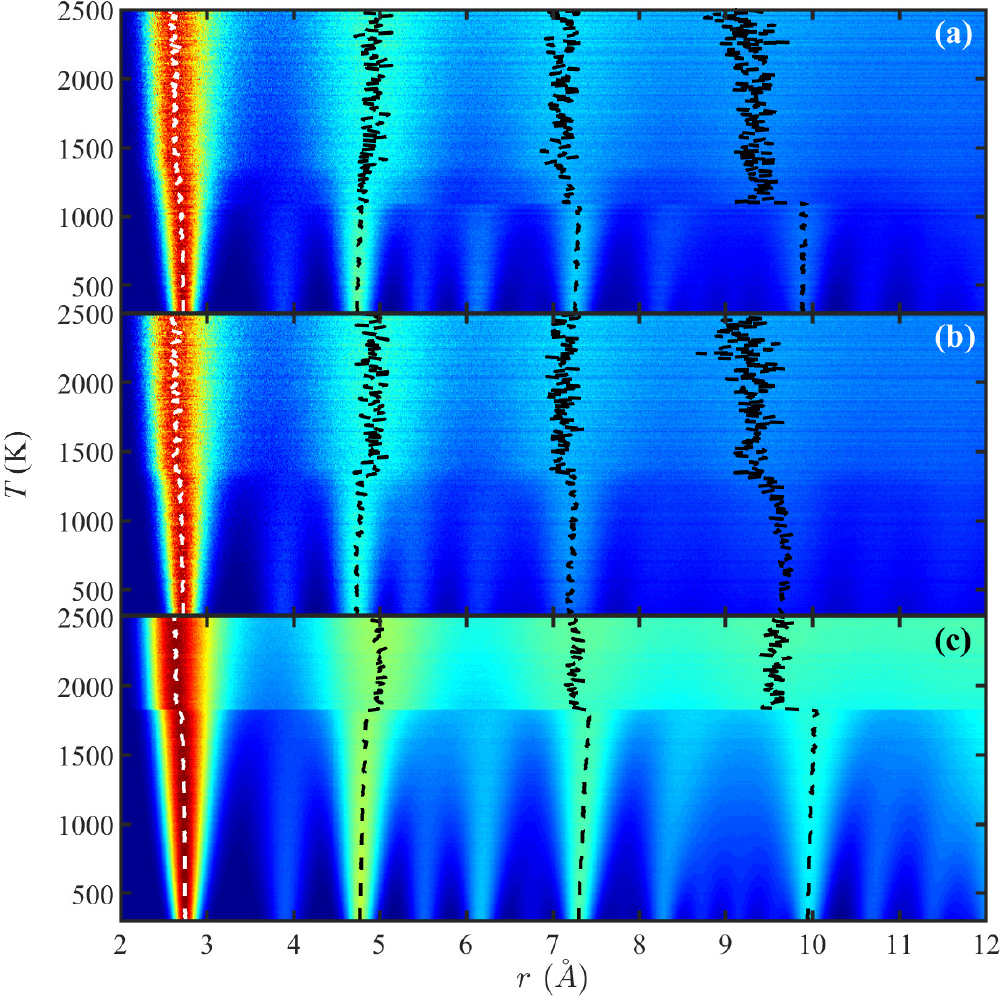}
    \caption{Variation of $G(r)$ with temperature for (a) 8-cub and (b) 8-ico clusters compared to that of (c) the bulk. The colorbar illustrates normalized $G(r)$ with main peaks indicated by dashed lines.}
    \label{fig:Gr}
\end{figure}

Figure~\ref{fig:Gr} depicts the variation of $G(r)$ with $T$ for 8-cub and 8-ico clusters using 12~{\AA} cutoff and $dr=0.01$~{\AA}. The figure also contains $G(r)$ of a bulk sample for comparison obtained for 32000 atoms with periodic boundary condition. In all cases, the $G(r)$ shows the same pattern for solid and liquid states in agreement with previously reported patterns \citep{Pan05}. The solid-state at the bottom of each figure can be interpreted from 4 main peaks indicated by dashed line corresponding to 4 shells (not to be confused with 4th nearest neighbors). The 1st shell is consisting of 12 equidistance atoms and thus presents a single peak while further shells contain more peaks due to their geometrical complexity. The difference between cub and ico cluster is limited to peaks at 3rd and 4th shells ($7<r<12$ \AA). The molten state at the top of each figure consisting of 4 broad peaks. It can be clearly seen that an increase in temperature causes broadening of peaks in the both solid and molten states. The melting can be interpreted as a step change in the position of inner shells peak (1st and 2nd dashed lines on the left). A $T_{mp}$ of about 1300~K can be detected for both figures in agreement with values obtained from caloric curves. In addition there is a step change at about 1100~K in 3rd and 4th dashed lines of the cub cluster (cf.~figure~\ref{fig:Gr}(a)). Since the step change is more evident in 4th dashed line corresponding to the 4th shell, \citet{Pan05} reported such a phenomenon as shell by shell melting. However, it is shown earlier that this step change belongs to cub to ico transformation. In the case of ico, the melting is associated with a smooth change in the peaks positions which is more evident in the 4th dashed line.

\subsubsection{Common neighbor analysis}

Recently, CNA has shown promising thanks to providing the possibility of distinction between allotropic transitions and melting process \citep{kateb2018}. The CNA identifies the crystal structure of each atom based on the concept of bond-orientational order parameter developed by \citet{Steinhardt83}. Briefly, the CNA determines local crystal structure based by decomposition of $G(r)$ into different angles \citep{Faken94}. Thus a twining grain boundary as the main difference of ico and cub cluster can be determined based on a slight angle difference between pairs of 1st nearest neighbors while it has the same number of 1st nearest neighbors as an fcc atom.

\begin{figure}
	\centering
    \includegraphics[width=1\linewidth]{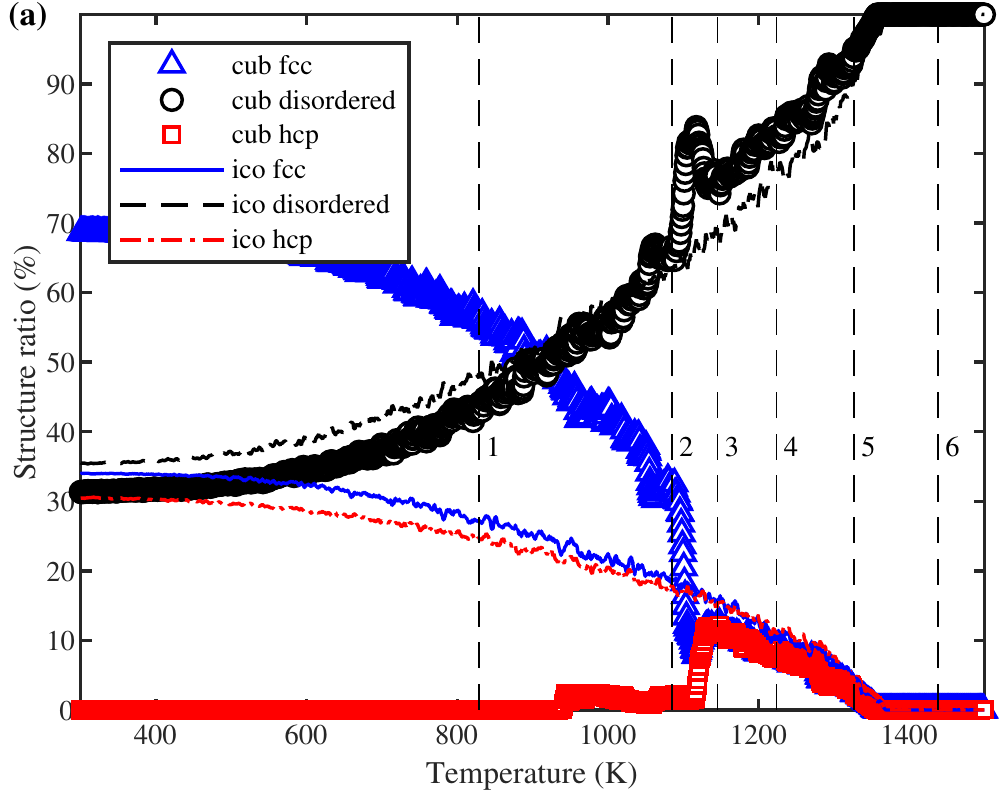}
	\includegraphics[width=1\linewidth]{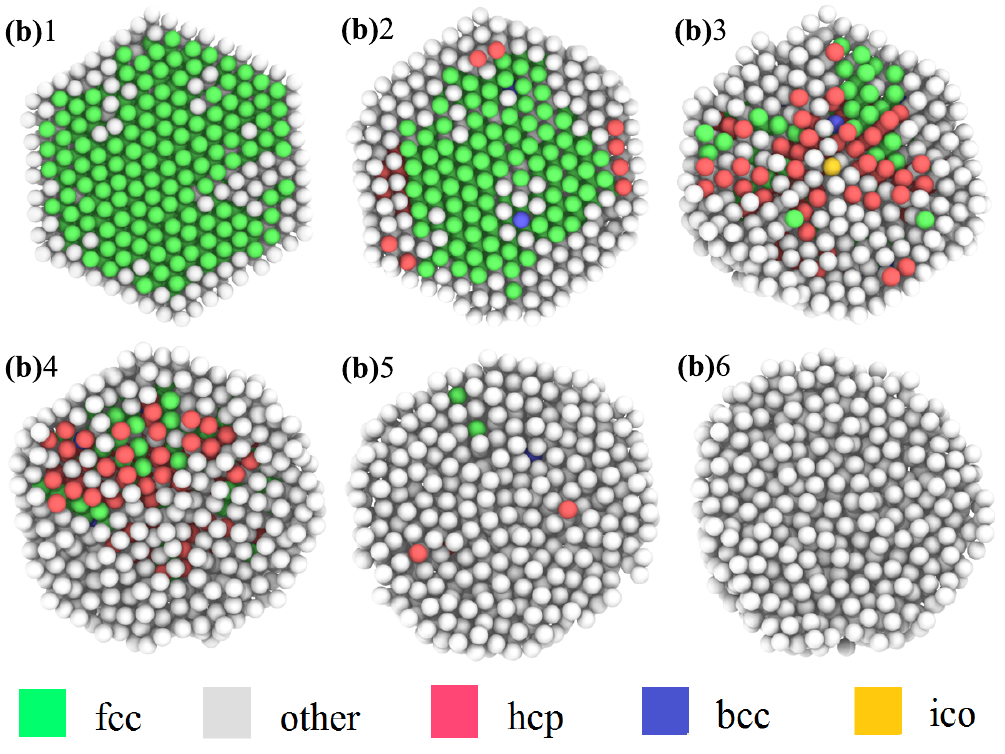}
    \caption{(a) Variation of fcc, hcp and disordered ratio with temperature for 8-cub and 8-ico clusters. (b) Snapshots of 8-cub cross section corresponding to the dashed lines 1 -- 6 in the (a).}
    \label{fig:cna}
\end{figure}

Figure \ref{fig:cna} presents the variation in the ratio of different structures obtained by CNA with temperature for 8-cub cluster. The figure also includes corresponding snapshots of the particle cross section at points 1 -- 6 indicated with vertical dashed lines. At 300~K, cub cluster (indicated by symbols) is made of 70\% fcc atoms and 30\% surface atoms characterized as disordered. While ico cluster indicated by lines consists of about 30\% of each fcc, hcp and surface atoms. As temperature increases, percentage of fcc and disordered atoms show a mirror change up to 1070~K. There is also a slight change in the percentage of hcp atoms as can be seen in snapshots (b)1 and (b)2. However, between 2 and 3 there is 24.6\% drop in fcc percentage and 19.1\% increase in disordered percentage. This is followed by 8.8\% drop in disordered atoms and nucleation of 9.3\% hcp and 4.3\% fcc atoms. As can be seen in snapshot b(3), there is higher ratio of disordered atoms at the bottom of cluster because of incomplete cub to ico transition which explains the slight difference between structure ratios of cub and ico cluster after the transition. 

\subsubsection{Potential energy distribution}

\begin{figure}
	\centering
    \includegraphics[width=1\linewidth]{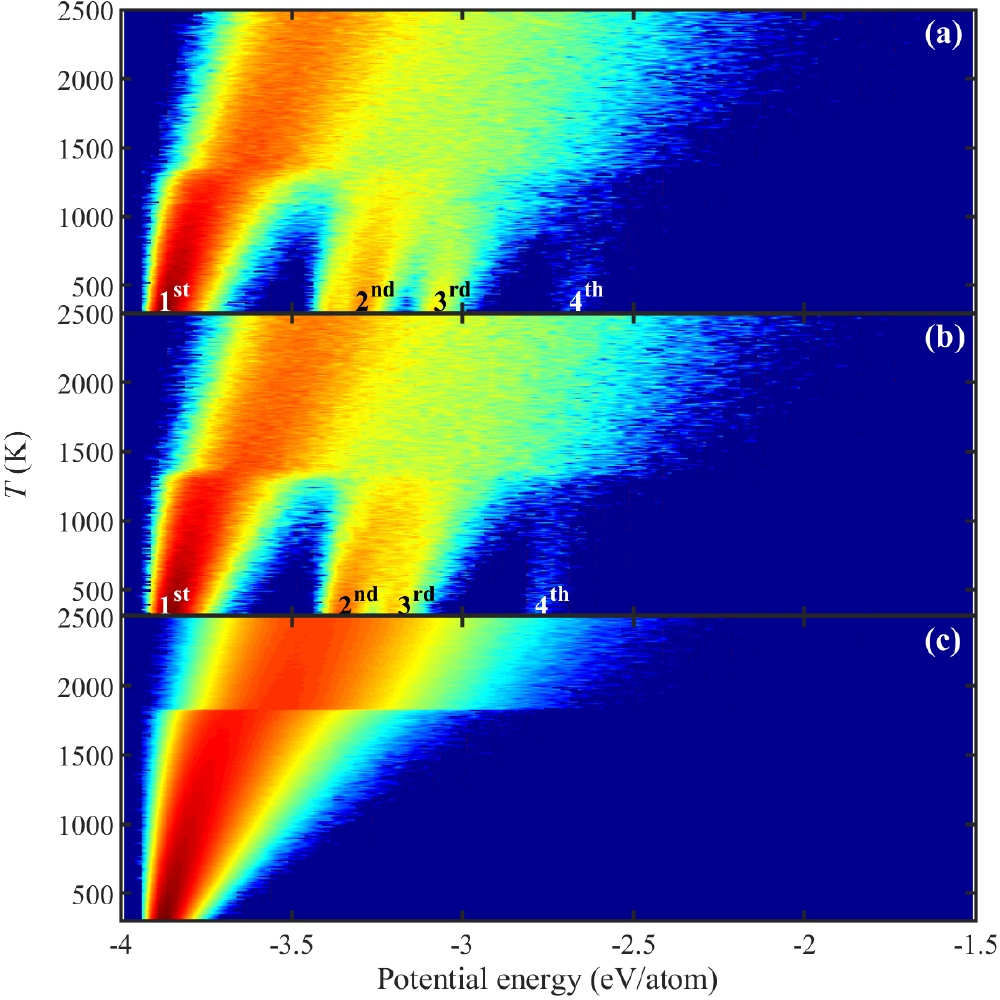}
    \caption{Variation of potential energy distribution with $T$ for (a) 8-cub and (b) 8-ico clusters in comparison with (c) the bulk. The colorbar indicates number of atoms possessing a specific energy in log scale.}
    \label{fig:Pe3d}
\end{figure}

A major difference between surface and interior atoms is their nearest neighbor which is reflected in the per atom potential energy. Thus it allows the study of surface related phenomena such as the surface melting more precisely. Figure \ref{fig:Pe3d} illustrates the variation of the potential energy distribution with $T$ for 8-cub and 8-ico clusters with the color bar being number of atoms that possess a specific energy in the log scale. The figure also contain the result of the bulk for comparison. Unlike $G(r)$, the features of the potential energy distribution show a clear difference between clusters and the bulk. At ambient temperature, a sharp peak is evident as indicated by 1$^{\rm st}$ corresponding to the interior atoms with potential energy about $E_c$ (3.935~eV/atom). There are also three minor peaks for both cub and ico with higher (more positive) potential energy which belong to surface atoms. Unlike 1$^{\rm st}$ peak, the position of 2$^{\rm nd}$ -- 4$^{\rm th}$ peaks are different in cub and ico. The dark blue region between 1$^{\rm st}$ and 2$^{\rm nd}$ peaks indicates the lack of atoms with intermediate potential energy and the fact that dividing a cluster into interior and surface atoms is correct assumption at low temperatures. Strictly speaking, however, the surface atoms in both clusters have to be divided into plane, edge and corner atoms corresponding to the 2$^{\rm nd}$ -- 4$^{\rm th}$ minor peaks, respectively. At elevated temperatures, the potential energy distribution and major peak become broader and there are no distinguishable minor peaks for both clusters. At this stage it is very hard to divide between inner and outer atoms and even at the particle-vacuum interface potential energy changes very smoothly.

As can be seen in both clusters, there is a step change in the major peak at about 1300~K corresponding to $T_{mp}$. It can be also seen that 3$^{\rm rd}$ cub peak vanishes at 1100~K while 2$^{\rm nd}$ and 4$^{\rm th}$ peaks showing negligible variation. One may think this is associated with the surface melting. However, it is expected that after surface melting the cluster corners disappear. While here the 4$^{\rm th}$ peak indicates the existence of corners above 1100~K. Comparing the 2$^{\rm nd}$ -- 4$^{\rm th}$ cub peaks with that of ico cluster, it appears that cub cluster present the same distribution as ico above 1100~K. This indicates that cub cluster transformed into ico below $T_{mp}$. Thus, the potential energy distribution can be utilized to determine shape of the cluster as well as melting and solid state transition temperatures. However, it is necessary to compare the potential energy distribution with the structure analysis result such as CNA to understand the exact origin of such variations in the potential energy distribution.

\subsubsection{Surface energy}
In MD simulation, the simplest way to determine the $\gamma_p$ is based on the slab model which is mainly defined at absolute zero where ab initio and MD are in close agreement \citep{Vitos98,Medasani09}.

\begin{equation}
	\gamma_p=\frac{U_p-N_tE_c}{4\pi R_p^2}
    \label{eq:slab}
\end{equation}
where the $U_p$ is the total potential energy of the cluster and $E_c$ is per atom cohesive energy of the bulk system. The denominator represents the cluster surface area for a spherical cluster of radius $R_p$. The later can be determined by Guinier formula \citep{Miao05}:

\begin{eqnarray}
	\label{eq:guinier}
	R_p=R_g\sqrt[]{\frac{5}{3}}+r_a\\
    R_g=\sqrt[]{\frac{1}{N}}\sum_i(r_i-r_{cm})^2
\end{eqnarray}
where $r_a$ is atomic radius and $R_g$ stands for cluster gyration radius with $r_i$ and $r_{cm}$ respectively being coordinates of atom $i$ and particle center of mass.

$U_p$ increases with the increase in $T$ (cf. figure~\ref{fig:caloric}(a)) and consequently Eq.~(\ref{eq:slab}) predicts increased $\gamma_p$ upon $T$ increment. This is in contradiction with Guggenheim-Katayama empirical formula \citep{Adam41}. Another statistical definition of $\gamma_p$ is based on the coordination number concept or so called \emph{broken bond model} \citep{Jiang08}.
\begin{equation}
	\gamma_b=(1-\frac{Z_s}{Z_b})\frac{E_c}{a_0}
    \label{eq:brokbond}
\end{equation}
where $a_0$ denotes the area of the two-dimensional unit cell of the solid. Eq.~(\ref{eq:brokbond}) states that $\gamma_b$ is directly proportional to the $E_c$. Assuming this stands correct for clusters, \citet{Jiang08} proposed the following size-dependent surface energy for the clusters.
\begin{equation}
	\frac{\gamma_p}{\gamma_b}=\frac{E_p}{E_c}=\left(1-\frac{1}{D/r_a-1}\right)\exp\left(-\frac{2E_c}{3RT_{sb}}\frac{1}{D/r_a-1}\right)
    \label{eq:jiang}
\end{equation}
with $E_p$ being cluster cohesive energy and $T_{sb}$ being the sublimation temperature of bulk material. One can adopt Eq.~(\ref{eq:jiang}) by substituting $E_p$ with $U$:

\begin{equation}
	\frac{\gamma_p}{\gamma_b}=\frac{U}{E_c}
    \label{eq:mine}
\end{equation}

The main advantage of Eq.~(\ref{eq:mine}) is the fact that it does not need the estimation of the cluster surface area.

\begin{figure}
	\centering
    \includegraphics[width=1\linewidth]{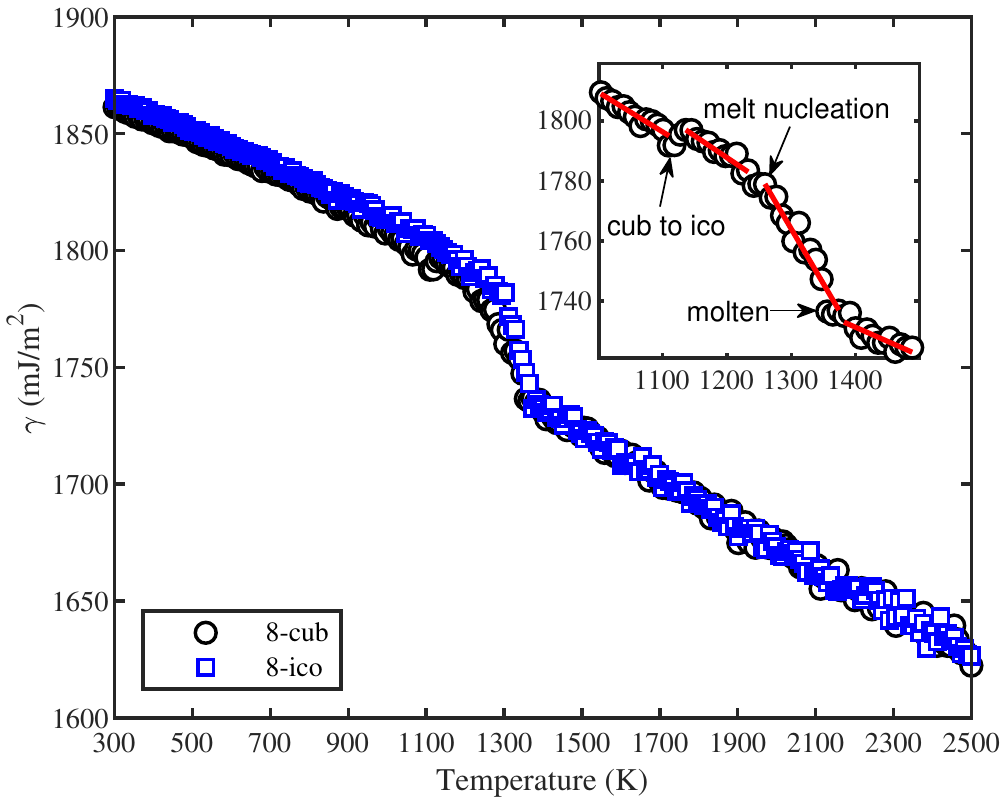}
    \caption{Variation of $\gamma_{p}$ with $T$ for 8-cub and 8-ico clusters. The inset magnifies transitions of 8-cub cluster.}
    \label{fig:GammavsT}
\end{figure}

The variation of $\gamma_p$ for 8-cub and 8-ico cluster with $T$ are shown in figure~\ref{fig:GammavsT}. It can be seen that Eq.~(\ref{eq:mine}) predicts the correct slop i.e. $\partial\gamma_p/\partial T<0$. The isotherm change in both curves indicate melting at about 1300~K. The figure inset shows the variation of $\gamma_p$ around the transitions of 8-cub cluster. The local minimum due to cub to ico transition can be clearly seen in the figure inset. We would like to remark that the slop of the curve ($\partial\gamma_p/\partial T$) remains the same before and after such a transition. However, ico structure is the most compact one and consequently presents smaller surface area and higher $\gamma_p$. Thus cub to ico transition appears as a jump in $\gamma_p$ while melting is associated with an expansion resulting in a drop in $\gamma_p$.

\subsection{Size-dependency}
\subsubsection{Size-dependent melting temperature}

Safaei \citep{Safaei10} developed a model considering the effects of the 1st nearest neighbors (NN) and the 2nd NN atomic interactions. An approximation of the formula without considering 2nd NN atomic interaction is as follows:

\begin{equation}
	\frac{T_{mp}}{T_{mb}}=1-\frac{N_s}{N_t}\left(1-q\frac{\epsilon_s}{\epsilon_i}\right),q=\frac{\bar{Z_s}}{Z_i}
    \label{eq:Tm}
\end{equation}
here $N_s$ stands for the number of surface atoms, $q$ is the coordination number ratio with $Z_i$ equal to 12 for interior atoms and $\bar{Z_s}$ as the average coordination numbers of surface atoms. The $\epsilon_i$ and $\epsilon_s$ respectively are bond energies of interior and surface atoms which the latter consists of cluster faces, edges and corners.

\begin{figure}
	\centering
    \includegraphics[width=1\linewidth]{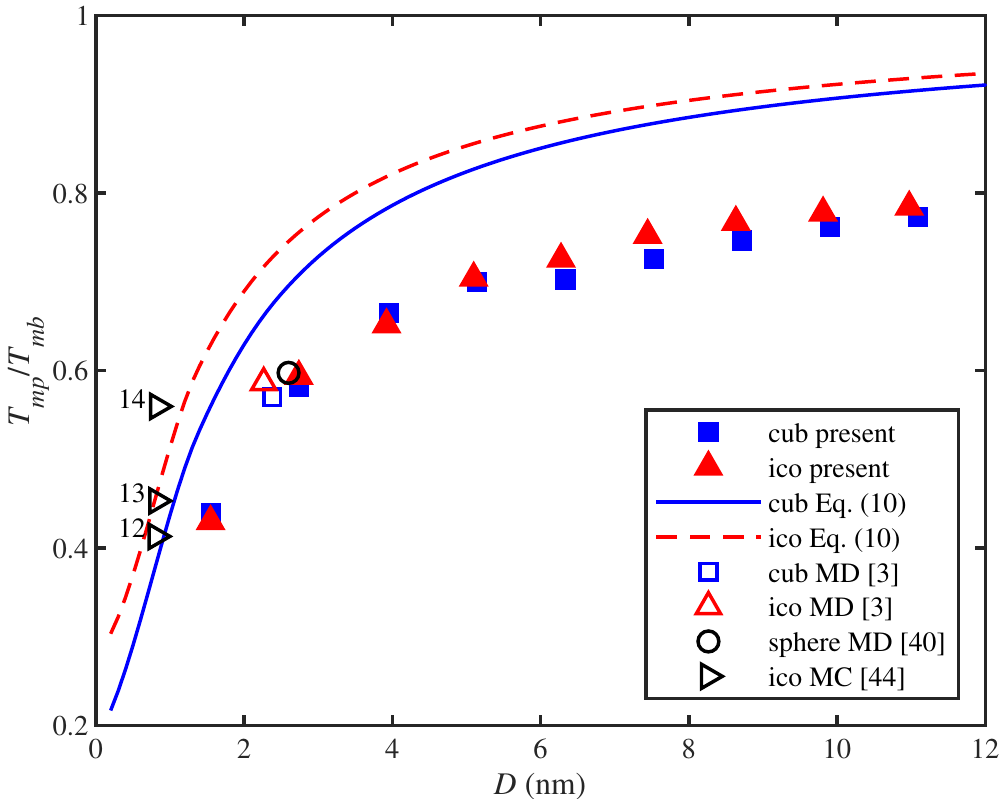}
    \caption{Dependence of the $T_{mp}$ on the size of particle obtained from $C_p$ in comparison with MD \citep{Pan05,Miao05} and MC simulation \citep{lee01} as well as Safaei model \citep{Safaei10}. All datasets are normalized to the experimental value of $T_{mb}=1825$~K \citep{wortis88}.}
    \label{fig:Tm}
\end{figure}

Figure \ref{fig:Tm} compares normalized $T_{mp}$ obtained from caloric curves in comparison with previous MD \citep{Pan05,Miao05} and MC \citep{lee01} simulations. The figure also contains values calculated from Eq.~(\ref{eq:Tm}) assuming that the bond strength for the surface and interior atoms to be equal. The melting points in the present study are in good agreement with previous MD result and lower than Safaei model \citep{Safaei10}. However, the model always predicts lower $T_{mp}$ for cub due to lower coordination number of (100) planes at the surface while ico surface is only made out of (111) planes with a higher coordination number. In this simulation, however, 2 -- 8-cub clusters are showing a different trend than higher shell numbers since 2 -- 4-cub are already transformed to ico during relaxation and 6 -- 8-cub transform to ico before melting. Thus, the first deficiency of the models is assuming a static crystallite without a crossover of different structures. Another difference is a smoother variation of $T_{mp}$ in the present result. These differences can also be explained by the static assumptions in the models. For instance, the $Z_i$ equal to 12 is not satisfied for sub-surface atoms at an elevated temperature when surface diffusion occurs or the ratio of fcc atoms drops as shown in figure \ref{fig:cna}. The MC result for clusters with 12 -- 14 atoms shows Pd$_{13}$ very well fitted with Safaei model and thus it is valid for full shell clusters. However, the model is unable to describe $T_{mp}$ of Pd$_{12}$ and Pd$_{14}$ clusters those reshaped to form a more stable cluster.

\subsubsection{Size-dependent melting enthalpy}

Attarian and Safaei \citep{Attarian08} proposed the following model for the size-dependent melting enthalpy of clusters.

\begin{multline}
	\frac{H_{mp}}{H_{mb}}=\left[1-2(1-q)\frac{D_0}{D+D_0}\right]\times\\
	\left\{1+\frac{3RT_{mb}}{2H_{mb}}\ln\left[1-2(1-q)\frac{D_0}{D+D_0}\right]\right\}
\end{multline}
%
with $D_0$ being a specific diameter which the entire atoms are located at the surface (i.e.\ $N_s=N_t$) and $H_{mb}$ is the bulk melting enthalpy.

\begin{figure}
	\centering
    \includegraphics[width=1\linewidth]{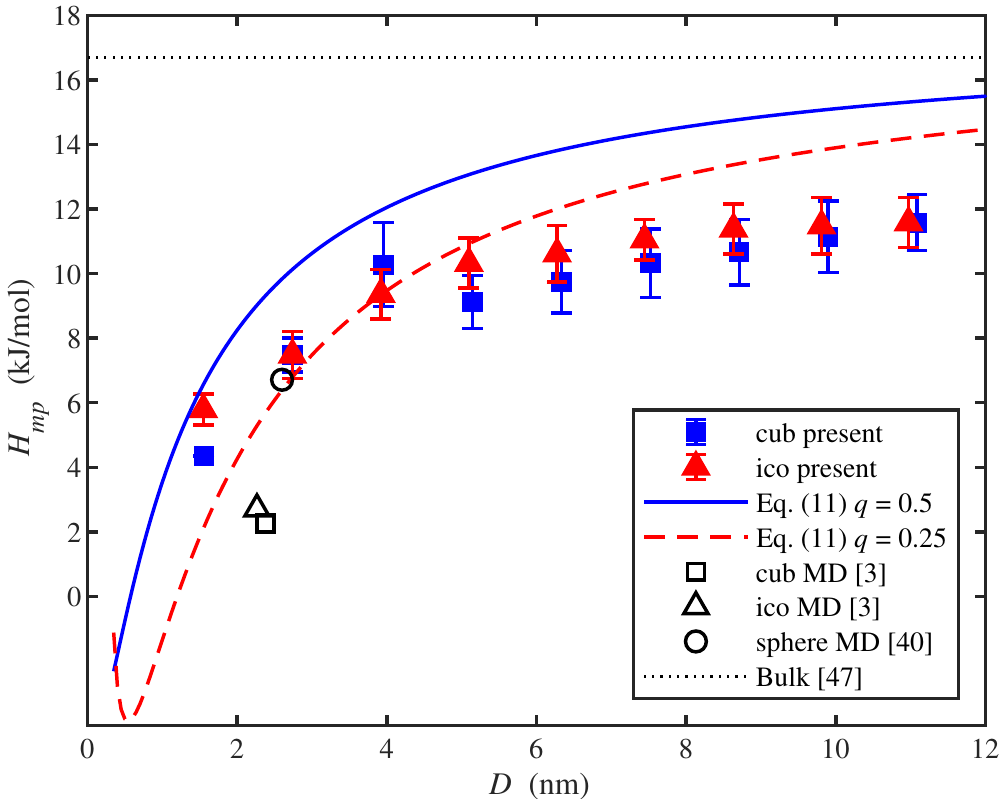}
    \caption{Variation of melting enthalpy with the diameter of Pd clusters, in comparison with previous MD results \citep{Pan05,Miao05} as well as Attarian and safaei \citep{Attarian08} model using $D_0=0.6712$~nm and the bulk value 16.7~kJ/mol \citep{Iida88}.}
    \label{fig:Hm}
\end{figure}

The variation of $H_{mp}$ with the particle size is shown in figure \ref{fig:Hm} in comparison with Attarian and Safaei \citep{Attarian08} model and previous MD results \citep{Pan05,Miao05}. However, the model shows a negligible difference for ico and cub and thus it is plotted for different $\bar{Z_s}$ of 6 and 3 corresponding to $q=0.5$ and 0.25, respectively. As can be seen, the model predicts an increase in the $H_{mp}$ with the particle size. However, the model determines $H_{mp}<0$ below 0.5754 and 1.1984 nm respectively for $\bar{Z_s}$ of 6 and 3 meaning that melting of smaller particles are exothermic and favorable. This failure is originated from crystalline basis of the model, which is not defined for a few atoms. Again, the results present a unified trend for ico clusters, while small cub clusters have a different trend than bigger ones. It can be seen that the result of \citet{Pan05} underestimates the $H_{mp}$ using SC potential.

\subsubsection{Size-dependent surface energy}

\begin{figure}
	\centering
    \includegraphics[width=1\linewidth]{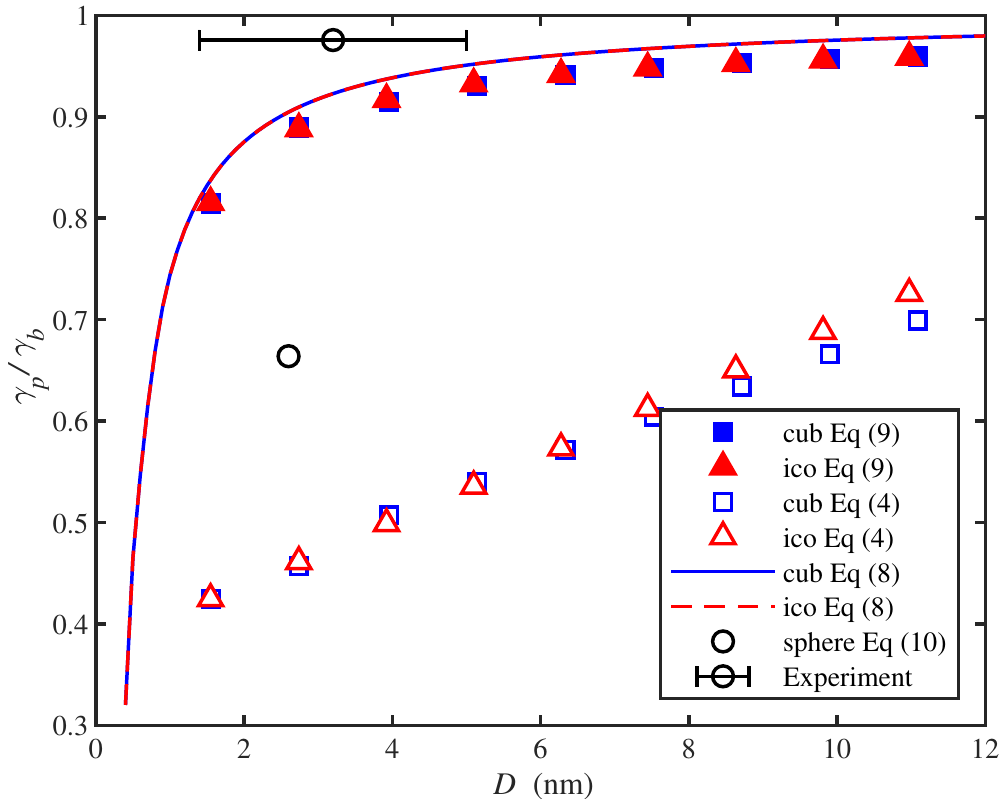}
    \caption{Variation of normalized $\gamma_p$ with the cluster size at 300~K using proposed, Eq.~(\ref{eq:mine}) and Eq.~(\ref{eq:slab}) in comparison with Jiang \citep{Jiang08} model and spherical cluster calculated by liquid drop model \citep{Miao05}. The experimental data was calculated in Ref \citep{Lu04} from surface stress of embedded clusters with weak particle matrix infractions \citep{Lamber95}. All datasets were normalized to the bulk value of 2050~mJ/m$^2$ \citep{Adam41}.}
    \label{fig:Gamma}
\end{figure}

Figure \ref{fig:Gamma} shows the normalized surface energy calculated using Eq.~(\ref{eq:slab}) and (\ref{eq:mine}), in comparison with Jiang model Eq.~(\ref{eq:jiang}), previous MD simulation \cite {Miao05} and experimental results \citep{Lu04,Lamber95}. The figure clearly shows the Eq.~(\ref{eq:slab}) (slab model) predicts a linear increase of $\gamma_p$ with the particle size. While Eq.~(\ref{eq:mine}) and Jiang model are showing a non-linear variation of $\gamma_p$ with the particle size in agreement with Tolman model \citep{Tolman49}, perturbation theory \citep{Samsonov99,Tayyebi14} and MC simulation \citep{Samsonov16}. This is very important since the latter predicts $\gamma_p$ = $\gamma_b$ for a infinity large particle while former results in values bigger than $\gamma_b$. Based on experimental data, Eq. \ref{eq:slab} and liquid drop model used in the calculation of spherical cluster underestimated the $\gamma_p$. It is worth noting that there is negligible difference between values calculated for cub and ico in all cases.
 
\section{Conclusion}

In conclusion, the stability and melting behavior of palladium clusters with 2 -- 20 shells of cub and ico structures studied using molecular dynamics simulation and EAM force field. The result shows small cub clusters are unstable at room temperature in agreement with experimental results. While cub clusters of intermediate size transform to ico at elevated temperatures and both cub and ico are stable up to the melting point for larger sizes. It is shown that $G(r)$ features are similar for the clusters and bulk. Thus $G(r)$ is not a surface sensitive method and needs to be compared with CNA to provide meaningful data on solid state transitions. While potential energy distribution gives different characteristics for the interior and surface atoms and even various sites at the surface i.e. plane, edge and corner. This allows the distinction between solid state transition and surface melting. Furthermore, utilizing $\gamma_p$ for detection of allotropic and melting transitions was compared to caloric curves i.e. potential energy and $C_p$. It is shown that cub to ico transition appears as a step change in potential energy which was neglected in the previous studies. While it appears as a minor peak in $C_p$ or a local minimum in $\gamma_p$ below melting point. However, the accuracy of $C_p$ depends on the number of samples averaged over. It is also shown that the well-known slab model predict the wrong trend for temperature and size dependency of clusters in contradiction with Guggenheim-Katayama and Tolman models, respectively. A simple relation was introduced based on available models that predict a correct trend for both size and temperature dependent $\gamma_p$.
\section*{Acknowledgments}

The Authors would like to thank Dr. Ebrahim Tayyebi and professor Hamid Modarress from Physical Chemistry groups respectively at University of Iceland and Amirkabir University of Technology for sharing their expertise in the surface energy calculation. This work is partially supported by University of Iceland research fund.

\section*{Data Availability}

The raw/processed data required to reproduce these findings cannot be shared at this time as the data also forms part of an ongoing study.


\bibliographystyle{model1-num-names}
\section*{\refname}
\bibliography{Ref.bib}




\end{document}